\documentclass[aps,prb,manuscript]{revtex4}
\usepackage{graphicx}
\usepackage{epsfig}
\usepackage{color}
\usepackage{bm}
\usepackage{float}

\begin{document}
\draft

\title{Valley- and spin-dependent quantum Hall states in bilayer silicene}

\author{Thi-Nga Do$^{1}\footnote{Corresponding author: {\em E-mail}: ngado@phys.ncku.edu.tw}$, Godfrey Gumbs$^{2,3}\footnote{Corresponding author: {\em E-mail}: ggumbs@hunter.cuny.edu}$, Po-Hsin Shih$^{1}$, Danhong Huang$^{4}$, Ming-Fa Lin$^{1}$}
\affiliation{$^{1}$Department of Physics, National Cheng Kung University, Taiwan 701\\
$^{2}$Department of Physics and Astronomy, Hunter College of the City University of New York,
695 Park Avenue, New York, New York 10065, USA \\
$^{3}$Donostia International Physics Center (DIPC), P de Manuel Lardizabal,
 4, 20018 San Sebastian, Basque Country, Spain\\
$^{4}$US Air Force Research Laboratory, Space Vehicles Directorate, Kirtland Air Force Base, New Mexico 87117, USA
}

\date{\today}

\begin{abstract}

The Hall conductivity $\sigma_{xy}$ of many condensed matter systems presents a step structure when a uniform perpendicular magnetic field is applied. We report the quantum Hall effect in buckled AB-bottom-top bilayer silicene and its robust dependence on the electronic valley and spin-orbit coupling. With the unique multi-valley electronic structure and the lack of spin degeneracy, the quantization of the Hall conductivity in this system is unlike the conventional sequence as reported for graphene. Furthermore, the conductivity plateaux take different step values for conduction ($2e^2/h$) and valence ($6e^2/h$) bands since their originating valleys present inequivalent degeneracy. We also report the emergence of fractions under significant effect of a uniform external electric field on the quantum Hall step structure by the separation of orbital distributions and the mixing of Landau levels from distinct valleys. The valley- and spin-dependent quantum Hall conductivity arises from the interplay of lattice geometry, atomic interaction, spin-orbit coupling, and external magnetic and electric fields.

\end{abstract}
\pacs{PACS:}
\maketitle

\section{Introduction}

The quantum Hall effect (QHE) for two-dimensional (2D) systems in the presence of a perpendicular magnetic field has attracted tremendous attention from both condensed matter physics theoreticians and experimentalists. The explanations for the QHE have been based on specific behaviors of the single-particle Landau level (LL) states arising under a magnetic field. The conductivity step structure could be understood in terms of the electrons filling Landau levels. The highly precise quantization of the integer QHE was demonstrated to be a consequence of gauge invariance and the existence of a mobility gap \cite{Lau}. More explicitly, they were explained by a topological invariant called the Chern number \cite{chern1,chern2}.

\medskip
\par

In this work, we report the interesting valley- and spin- dependent QHE of buckled bilayer silicene based on a robust connection with the lattice geometry. We accomplish this by numerically calculating the Hall conductivity for AB-bottom-top (bt) stacking when a perpendicular magnetic field is applied. The special lattice configuration of the system induces complex interlayer atomic interactions and significant spin-orbit couplings (SOCs). We demonstrate below that the interplay between these intrinsic characteristics and an external perpendicular magnetic field leads to the extraordinary quantization of LLs. Consequently, the quantum Hall conductivity (QHC) exhibits integer step structures with different sequences, depending on the initiating valley of LLs. Such unusual quantization of the QHE could be remarkably enriched by the application of a perpendicular electric field, specifically the emergence of fractions by the separation of orbital distributions and the odd plateau sequence due to mixing of Landau levels from distinct valleys. These diverse magneto-transport properties have never been reported in the literature for condensed matter systems \cite{ mono, ABA, ABC, twist, BP}. This demonstrates the essential role played by the crystal structure in the QHE.  Our theoretical predictions open the door to new possibilities in understanding the nature of quantum Hall conductivity quantization in 2D materials.

\medskip
\par

\begin{figure}[b]
\centering
{\includegraphics[width=0.5\linewidth]{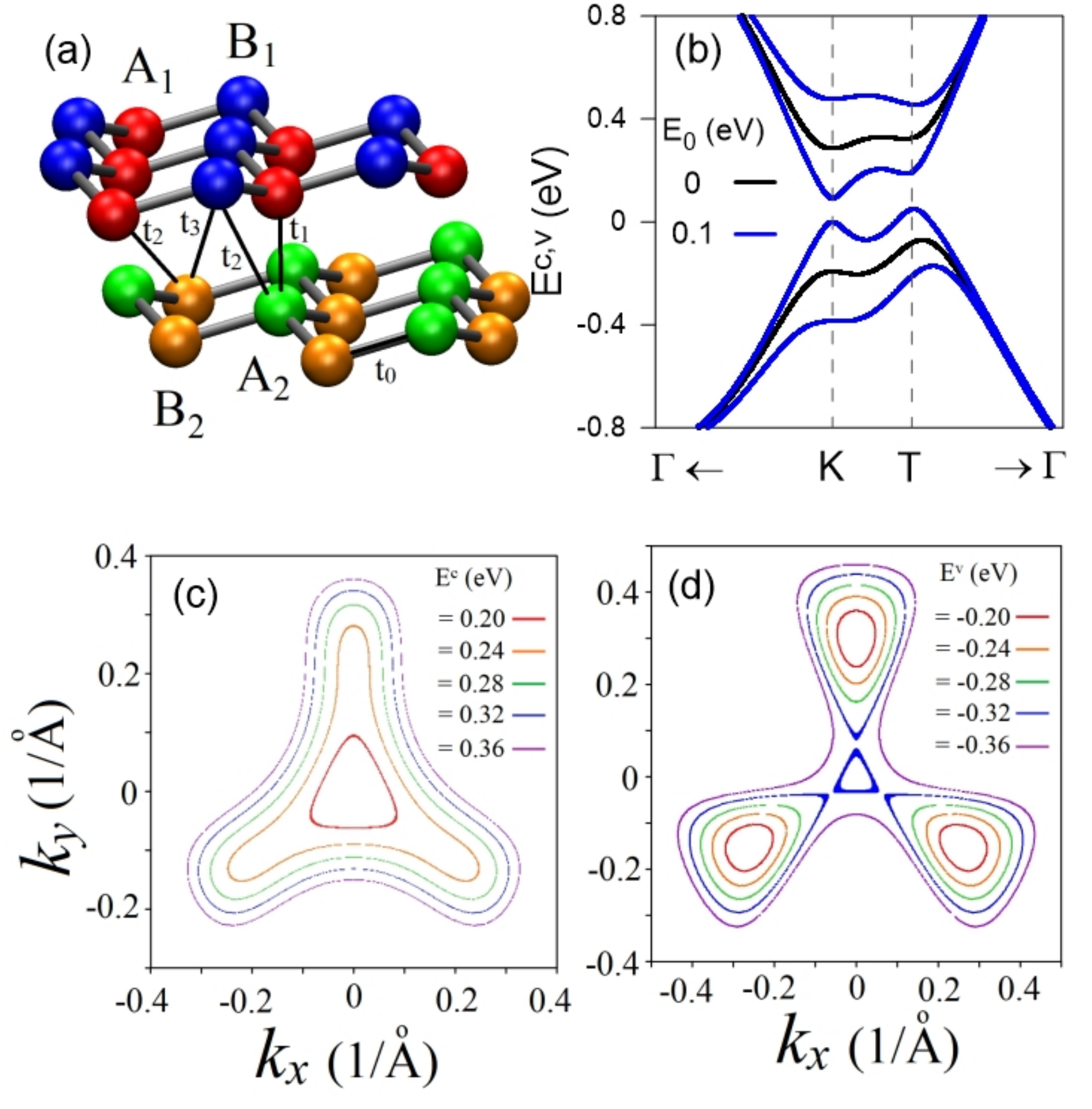}}
\caption{(color online)  (a) The side view of the atomic structure for AB-bt bilayer silicene and (b) its band structures for zero and finite perpendicular electric fields. The constant-energy diagrams are presented for (c) the conduction and (d) the  valence subbands.}
\label{Fig1}
\end{figure}

Silicene has been successfully synthesized on different substrates \cite{Si, Si1, Si2, Si3}. This novel 2D material has also been the subject of numerous theoretical studies due to its exotic electronic structure and promising applications in silicon-based electronic technology \cite{Si4, Si5, Si6, Si7, Si8}. Layered silicene was predicted to show interesting physical features beyond its monolayer counterparts, such as topological and superconducting properties \cite{Si7, Si8}. So far, high-angle annular dark field scanning transmission electron microscopy has verified bilayer silicene with various stacking positions of the A and B atoms of the underlying lattice (see Fig.\ \ref{Fig1}(a)) and buckling orderings \cite{Si3}. Among four types of bilayer systems, AB-bt has proven to be the most stable stacking \cite{Si3}. It possesses a sizable band gap between the oscillatory energy dispersion which yields high potential for semiconductor applications and interesting magnetic quantization behaviors \cite{Si3, Si4, prb1}. Here, we show that the combined influence of lattice geometry, atomic interactions, SOCs, and external magnetic and electric fields on the Hall conductivity of this system gives rise to the integer and fractional QHE with peculiar sequences of conductivity plateaux.

\medskip
\par

\section{Method}

Within linear response theory, we calculated the Hall conductivity from the dynamic Kubo formula in the form of \cite{Kubo}

\begin{eqnarray}
\sigma_{xy} = \frac {ie^2 \hbar} {S}
&\sum_{\alpha} \sum_{\beta \neq \alpha} (f_{\alpha} - f_{\beta})
\frac {\langle \alpha  |\mathbf{\dot{u}}_{x}| \beta\rangle  \langle \beta |\mathbf{\dot{u}}_{y}|\alpha \rangle} {(E_{\alpha}-E_{\beta})^2 + \Gamma ^2} \ .
\label{eqn:1}
\end{eqnarray}
The Landau energies ($E_{\alpha,\beta}$) and wave functions ($|\alpha, \beta \rangle$) of the initial and final states in the inter-Landau level (LL) transitions are evaluated from the tight-binding Hamiltonian \cite{prb1}, which can be written as

\begin{eqnarray}
\nonumber
H & = &\sum_{m,l}(\epsilon_m^l+U_m^l)c_{m \alpha}^{\dagger l}c_{m \alpha}^{l}
+\sum_{ m,j  , \alpha, l, l^{\prime}} t_{mj}^{ll^{\prime}} c_{m \alpha}^{\dagger l}
c_{j \alpha}^{l^{\prime}}\\
\nonumber
&+&  \frac {i} {3\sqrt{3}} \sum_{ \langle \langle m,j \rangle \rangle, \alpha, \beta, l}
\lambda^{SOC}_{l} \gamma_l v_{mj} c_{m\alpha}^{\dagger l} \sigma_{\alpha\beta}^{z} c_{j\beta}^{l}\\
&-& \frac{2i}{3} \sum_{\langle\langle m,j \rangle \rangle, \alpha, \beta, l} \lambda^{R}_{l} \gamma_l u_{mj}
c_{m\alpha}^{\dagger l} (\vec{\sigma} \times \hat{d}_{mj})_{\alpha\beta}^{z} c_{j\beta}^{l}\ .
\label{eqn:2}
\end{eqnarray}
Here, $c_{m\alpha}^{l}$ ($c_{m\alpha}^{\dagger l}$) is the annihilation (creation) operator, $\epsilon_m^l (A^l,B^l)$ the site energies, and $U_m^{l} (A^l,B^l)$ the height-dependent Coulomb potential energies. The hopping terms, $t_{mj}^{ll^{\prime}}$, and SOCs, $\lambda_{1,2}^{SOC}$ and $\lambda_{1,2}^{R}$, are chosen in order to reproduce the band structure from the first-principle result \cite{Si5}. The matrix elements of velocity operators, $\mathbf{\dot{u}}_{x, y}$, directly determine the available inter-LL transitions. They can be efficiently computed from the gradient approximation, as successfully done for many condensed matter systems, such as carbon-related materials \cite{gradientapp}.

\medskip
\par

\section{Results and Discussion}

It is worthwhile noting that AB-bt bilayer silicene possesses a buckled lattice arrangement and a slightly mixed sp$^2$-sp$^3$ chemical bonding. Furthermore, the two silicene sheets have opposite buckled orderings (Fig. \ref{Fig1}(a)), leading to significant inter-layer atomic interactions and layer-dependence of the SOCs which generate its special band structure. There are two pairs of energy bands, for which the low-lying pair plays the main role in many unusual physical properties of the system. The low-energy conduction and valence bands are mainly determined by the $3p_{z}$ orbitals, exhibiting a sizable band gap and the evidently  asymmetric behavior near the Fermi energy $E_F=0$, as presented in Fig. \ref{Fig1}(b). The conduction band originates from the $\bf{K}$ ($\bf{K}^{\prime}$) valley while the valence band starts from a specific point midway between $\bf{K}$ ($\bf{K}^{\prime}$) and $\bf{\Gamma}$, which we refer to as the $\bf{T}$ ($\bf{T}^{\prime}$) valley. The main features of the energy dispersion are remarkably modified by an external electric field, referreing to the blue curves in Fig. \ref{Fig1}(b). In particular, the field reduces the band gap, separates the orbital distributions for each band, and enhances the oscillation at each valley. In general, both conduction and valence bands present peculiar oscillatory and strong anisotropic properties, clearly demonstrated by the constant-energy contours in Figs. \ref{Fig1}(c) and \ref{Fig1}(d). The conduction and valence electronic states nearer $E_F$ (the red curves), which are respectively around the $\bf{K}$ and $\bf{T}$ valleys, show  non-circular energy contours. The anisotropy becomes more obvious for higher conduction and deeper valence energy bands. The conduction (valence) states only come to exist near the $\bf{T}$ ($\bf{K}$) valley for sufficiently high (deep) energy. The unique features of the band structure and its sensitive behavior in an electric field give rise to the extraordinary QHE quantization, as we discuss below.

\begin{figure}[h]
\centering
{\includegraphics[width=0.5\linewidth]{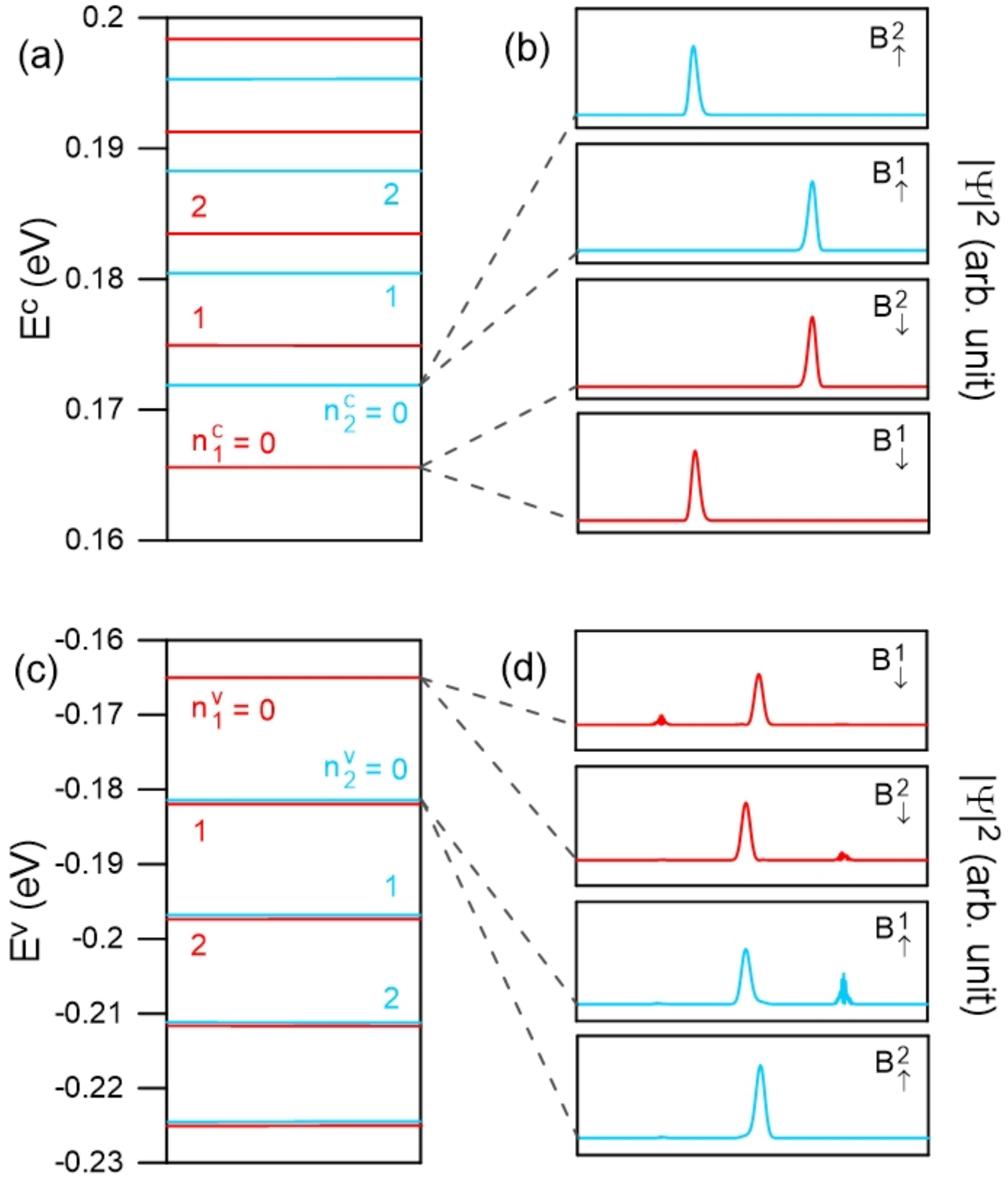}}
\caption{(color online) The (a) conduction Landau level energy spectrum and (b) wave functions on the dominant sublattices of $n^c_1 = 0$ LLs for $B_0$ = 40 T. The blue and red curves represent, respectively, the spin-up and spin-down dominated LLs. Similar plots for the valence band are presented in (c) and (d). }
\label{Fig2}
\end{figure}
The distinctive lattice geometry produces eight non-equivalent sublattices of four orbitals with two spin states, which play a decisive role in the unconventional QHE in the presence of a magnetic field. The remarkable difference between the A and B atoms due to their chemical environment raises the notion of {\em dominant\/} sublattices.  Our numerical calculations show  that the spatial distribution of quantized LLs originating from the low-lying energy bands have much larger amplitude on the B sublattices than the A ones. Quite the opposite is true for LLs quantized from the outer pair. Consequently, the B sublattices mainly contribute to the transitions between LLs near $E_F$. Both the conduction and valence LLs can be classified into four separate subgroups based on the behavior of their spatial distribution on the {\em dominant\/} B sublattices with two spin states. For each subgroup, the wave functions are dominated by one sublattice among $B_{1\uparrow}$, $B_{1\downarrow}$, $B_{2\uparrow}$ and $B_{2\downarrow}$, as illustrated for the four $n^c$ = 0 and four $n^v$ = 0 LLs in Figs. 2(a)-(d). Moreover, LLs are split into spin-up (blue) and spin-down (red) states because of the significant SOC in bilayer silicene.

\begin{figure}[h]
\centering
{\includegraphics[width=0.5\linewidth]{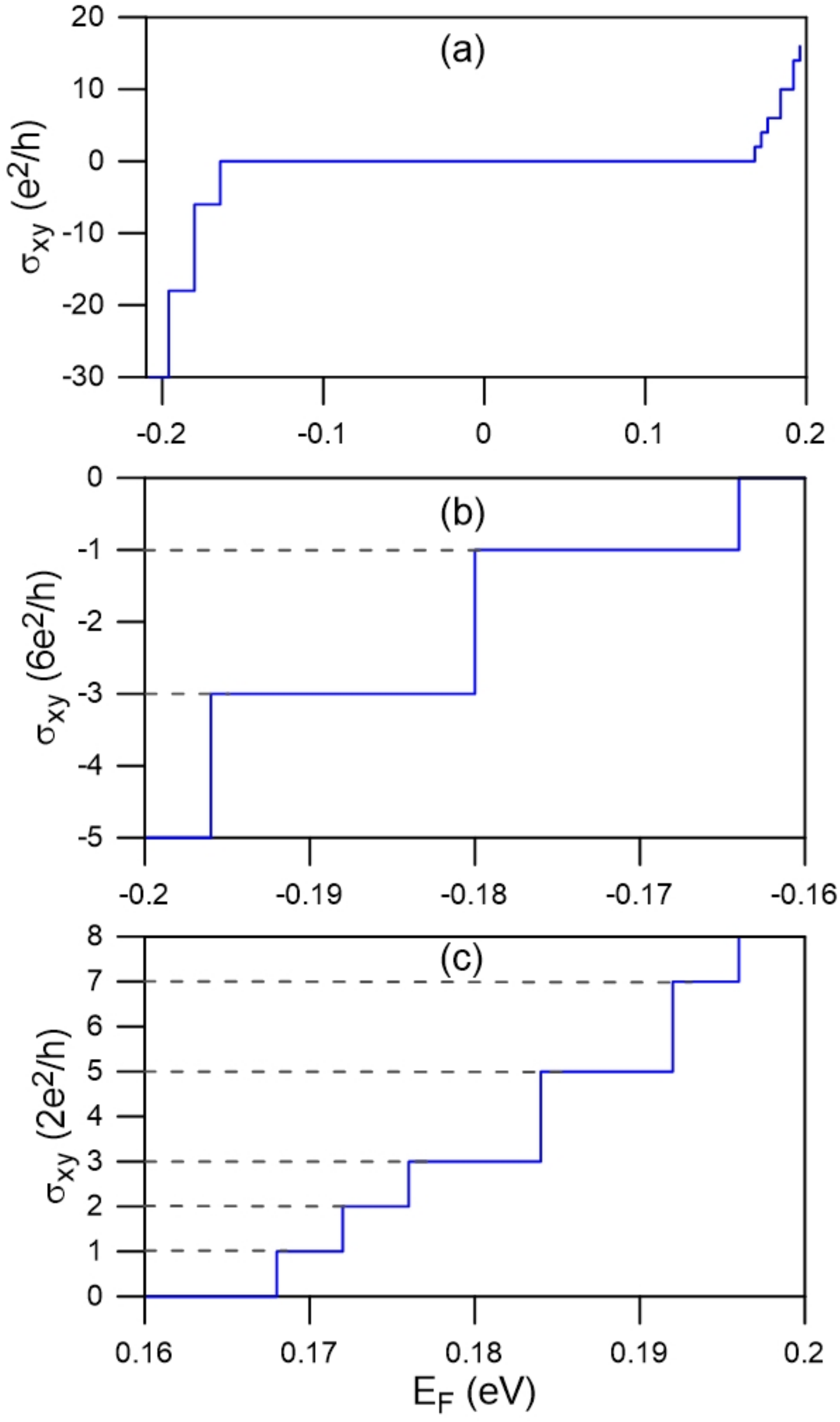}}
\caption{(color online) The (a) Fermi energy-dependent Hall conductivities for bilayer AB-bt silicene when a perpendicular magnetic field $B_z$ = 40 T is applied. A closer examination of the valence and conduction band conductivities  is shown in (b) and (c), respectively.}
\label{Fig3}
\end{figure}

The degeneracy of quantized LLs is of special importance in accounting for the height of quantum Hall steps. Here, we show that it is associated with the unique valley and spin properties. Similar to other honeycomb lattice systems such as graphene \cite{degeneracy1, degeneracy2, graphene1}, LLs originating from the $\bf{K}$ and $\bf{K}^{\prime}$ valleys are degenerate.  This is also true for LLs quantized from the $\bf{T}$ and $\bf{T}^{\prime}$ valleys due to special valley symmetry characteristics. Interestingly, the number of $\bf{T}$ and $\bf{T}^{\prime}$ points is thrice that for the $\bf{K}$ and $\bf{K}^{\prime}$ points in each first Brillouin zone. Therefore, the spin-split LLs at $\bf{T}$ ($\bf{T}^{\prime}$) valleys are six-fold degenerate while they are doubly degenerate for the $\bf{K}$ ($\bf{K}^{\prime}$) valleys. Moreover, the low-lying LLs are initiated from either the $\bf{K}$ ($\bf{K}^{\prime}$) point for the conduction band or $\bf{T}$ ($\bf{T}^{\prime}$) for the valence spectrum. These peculiar quantization phenomena are directly reflected in the quantum Hall conductivity, as we discuss in detail below.

\begin{figure}[h]
\centering
{\includegraphics[width=0.5\linewidth]{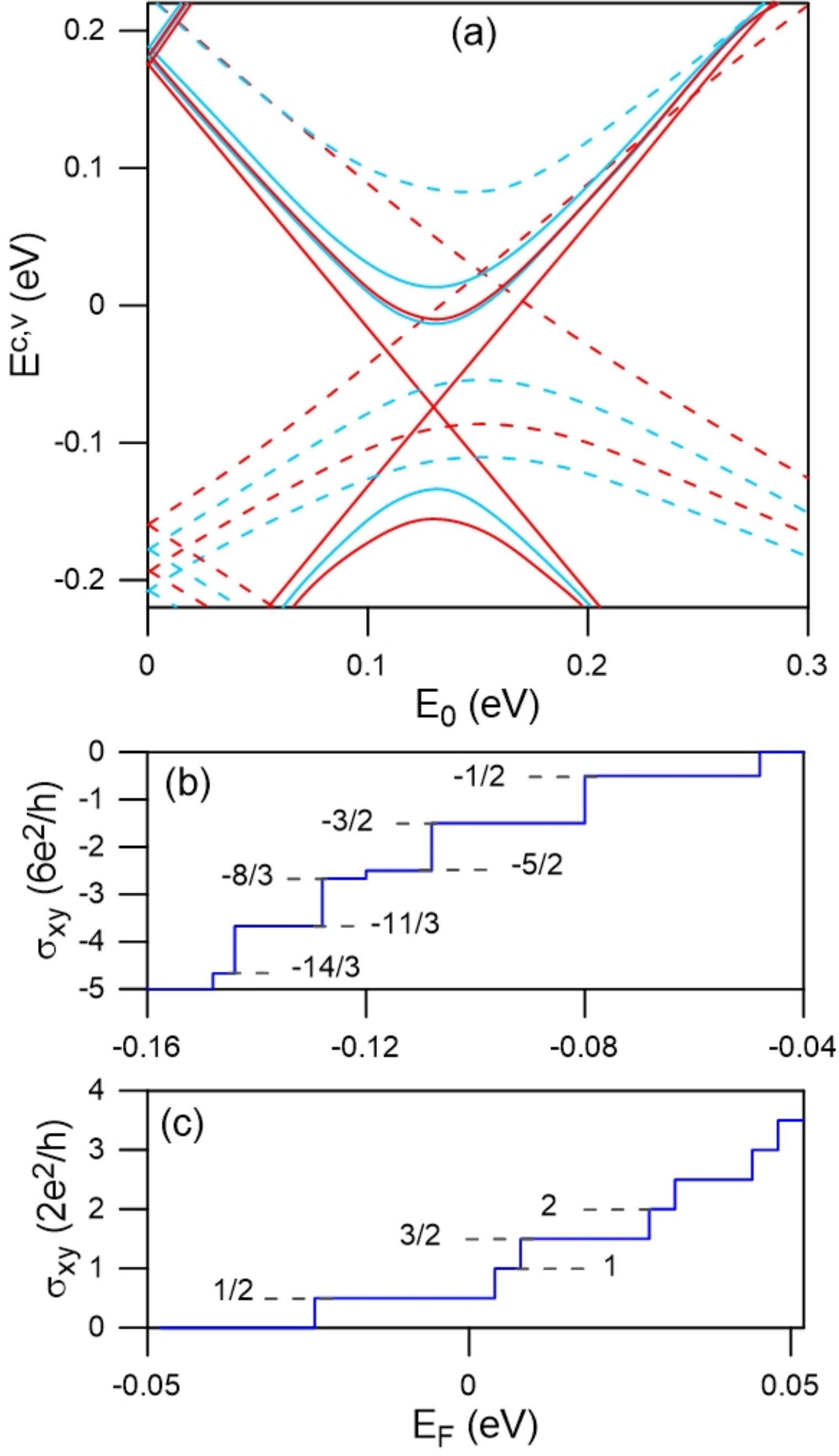}}
\caption{(color online) The (a) dependence of LL energies in a uniform perpendicular electric field for $B_z$ = 40 T. The solid and dashed curves denote the LLs originating from the $\bf{K}$ and $\bf{T}$ valleys, respectively. (b) and (c) show the Fermi energy-dependent Hall conductivities for the valence and conduction spectra for a finite electric field $E_0$ = 100 meV. }
\label{Fig4}
\end{figure}

We now turn our attention to an interpretation of the QHE of AB-bt bilayer silicene in the presence of a magnetic field. The Fermi energy-dependent QHC is quantized as integer multiples of $e^2/h$ for both the low-lying valence ($\sigma_{xy}=6e^2/h$) and conduction ($\sigma_{xy}=2e^2/h$) LLs, as depicted in Figs. \ref{Fig3}(a) through \ref{Fig3}(c). The step structure could be understood through inter-LL transitions, as clearly indicated in the formula of Hall conductivity in Eq. (\ref{eqn:1}) in terms of the velocity matrix elements. During the variation of the Fermi level ($E_F$), a plateau appears whenever one LL becomes occupied. On the other hand, the QHC steps of $6e^2/h$ and $2e^2/h$ are associated with the six-fold degenerate valence LLs and doubly degenerate conduction ones, respectively. As an exception, the overlap of valence LLs with $\Delta n$ = 1 leads to several double steps where the two plateaus differ by 12 instead of 6 in the unit of $e^2/h$. The QHC steps in bilayer silicene is in great contrast with the conventional $4e^2/h$ in bilayer graphene due to the $\pm p\hat{z}$ symmetry and spin degree of freedom induced four-fold degenerate LLs \cite{graphene1}.

\medskip
\par

The Hall conductivity at low Fermi energy yields both integer and fractional steps when a uniform perpendicular electric field is applied to the system. The principal reason is that, a finite $E_0$ dramatically changes the main features of the LLs, especially the degenerate degree of freedom and the originated electronic valleys.  As a matter of fact, the field induces separation of LLs dominated by the $B^1$ and $B^2$ sublattices, as shown in Fig. \ref{Fig4}(a) for the $E_0$-dependent LLs. It is crucial to note that, the fractional QHE we refer to is in the sense that we take the unit to be $ge^2/h$,  where $g$ is the LL degeneracy degree of freedom.
We observed the unconventional QHC quantization sequences of $(m-1/2) 6e^2/h$ and $(m^{\prime}/2) 2e^2/h$ ($m$ = 0, -1, -2,...; $m^{\prime}$ = 1, 2, 3,...) for valence and conduction spectra, respectively, referring to Figs. \ref{Fig4}(b) and \ref{Fig4}(c). Moreover, the emergence of valence LLs at the  $\bf{K}$ valley and conduction LLs at the $\bf{T}$ valley in the low energy range enriches the conductivity spectra with more plateaus, as shown in Fig. \ref{Fig4}(b) for the valence band. This creates an unique sequence of QHC with the steps at -8/3, -11/3, -14/3, -15/3... which has never been reported in the literature for any other 2D materials.   Critical $E_0$'s may significantly change the characteristics of the system, as revealed in the $E_0$-dependent LL spectrum. As a result, the existence of zero conductivity is strongly dependent on the $E_0$ strength through the band gap closing and opening behaviors. The crossing phenomenon of the conduction and valence LLs at $\bf{K}$ ($E_0$ = 130 meV) and $\bf{T}$ ($E_0$ = 153 meV) valleys leads to the absence of zero conductivity, similar in behavior to that in monolayer graphene \cite{monographene}.  The $E_0$-controlled QHC in AB-bt bilayer silicene may have high potential in  $Si$-based electronic device applications.

\medskip
\par

\section{Concluding Remarks}

In summary, we have investigated the QHE in bilayer silicene and explained its nature based on its intrinsic material properties as well as the influence of an external field. The lattice geometry, atomic interaction, SOCs, and external magnetic and electric fields are responsible for the extraordinary integer and fractional QHC for the AB-bt system. The quantization of QHC can be manipulated by an external electric field through the separation of orbital distributions and the mixing of Landau levels from distinct valleys. Our discovery opens up a possible new  physical mechanism which serves to account for the unusual sequence of the step structure. The calculated results, obtained from the efficient combination of the Kubo formula and the generalized tight-binding model, can be useful for comparison in transport experiments with silicene.

\medskip
\par

\begin{acknowledgements}
The authors thank the Ministry of Science and Technology of Taiwan (R.O.C.) for financial support under Grant \# MOST 105-2112-M-017-002-MY2. DH would also like to acknowledge the support from the Laboratory University Collaboration Initiative (LUCI) program and from the Air Force Office of Scientific Research (AFOSR).
\end{acknowledgements}

\end{document}